\documentclass{WileyMSP-template}
\usepackage{graphicx}
\usepackage{dcolumn}
\usepackage{bm}  
\usepackage{amssymb}  
\usepackage{upgreek}
\usepackage{amsmath}
\usepackage{verbatim}
\usepackage[T1]{fontenc}
\usepackage[dvipsnames]{xcolor}
\usepackage{lmodern}
\usepackage[utf8]{inputenc}
\usepackage[english]{babel}
\usepackage{hyperref}
\hypersetup{colorlinks=true,citecolor={black},linkcolor={black},urlcolor={black}}
\usepackage{soul}

\usepackage[normalem]{ulem}
\RequirePackage[normalem]{ulem} 
\RequirePackage{color}\definecolor{RED}{rgb}{1,0,0}\definecolor{BLUE}{rgb}{0,0,1} 
\providecommand{\DIFaddbegin}{} 
\providecommand{\DIFaddend}{} 
\providecommand{\DIFdelbegin}{} 
\providecommand{\DIFdelend}{} 
\providecommand{\DIFaddbeginFL}{} 
\providecommand{\DIFaddendFL}{} 
\providecommand{\DIFdelbeginFL}{} 
\providecommand{\DIFdelendFL}{} 
\newcommand{\DIFscaledelfig}{0.5}
\RequirePackage{settobox} 
\RequirePackage{letltxmacro} 
\newsavebox{\DIFdelgraphicsbox} 
\newlength{\DIFdelgraphicswidth} 
\newlength{\DIFdelgraphicsheight} 
\LetLtxMacro{\DIFOincludegraphics}{\includegraphics} 
\newcommand{\DIFaddincludegraphics}[2][]{{\color{blue}\fbox{\DIFOincludegraphics[#1]{#2}}}} 
\newcommand{\DIFdelincludegraphics}[2][]{
\sbox{\DIFdelgraphicsbox}{\DIFOincludegraphics[#1]{#2}}
\settoboxwidth{\DIFdelgraphicswidth}{\DIFdelgraphicsbox} 
\settoboxtotalheight{\DIFdelgraphicsheight}{\DIFdelgraphicsbox} 
\scalebox{\DIFscaledelfig}{
\parbox[b]{\DIFdelgraphicswidth}{\usebox{\DIFdelgraphicsbox}\\[-\baselineskip] \rule{\DIFdelgraphicswidth}{0em}}\llap{\resizebox{\DIFdelgraphicswidth}{\DIFdelgraphicsheight}{
\setlength{\unitlength}{\DIFdelgraphicswidth}
\begin{picture}(1,1)
\thicklines\linethickness{2pt} 
{\color[rgb]{1,0,0}\put(0,0){\framebox(1,1){}}}
{\color[rgb]{1,0,0}\put(0,0){\line( 1,1){1}}}
{\color[rgb]{1,0,0}\put(0,1){\line(1,-1){1}}}
\end{picture}
}\hspace*{3pt}}} 
} 
\LetLtxMacro{\DIFOaddbegin}{\DIFaddbegin} 
\LetLtxMacro{\DIFOaddend}{\DIFaddend} 
\LetLtxMacro{\DIFOdelbegin}{\DIFdelbegin} 
\LetLtxMacro{\DIFOdelend}{\DIFdelend} 
\DeclareRobustCommand{\DIFaddbegin}{\DIFOaddbegin \let\includegraphics\DIFaddincludegraphics} 
\DeclareRobustCommand{\DIFaddend}{\DIFOaddend \let\includegraphics\DIFOincludegraphics} 
\DeclareRobustCommand{\DIFdelbegin}{\DIFOdelbegin \let\includegraphics\DIFdelincludegraphics} 
\DeclareRobustCommand{\DIFdelend}{\DIFOaddend \let\includegraphics\DIFOincludegraphics} 
\LetLtxMacro{\DIFOaddbeginFL}{\DIFaddbeginFL} 
\LetLtxMacro{\DIFOaddendFL}{\DIFaddendFL} 
\LetLtxMacro{\DIFOdelbeginFL}{\DIFdelbeginFL} 
\LetLtxMacro{\DIFOdelendFL}{\DIFdelendFL} 
\DeclareRobustCommand{\DIFaddbeginFL}{\DIFOaddbeginFL \let\includegraphics\DIFaddincludegraphics} 
\DeclareRobustCommand{\DIFaddendFL}{\DIFOaddendFL \let\includegraphics\DIFOincludegraphics} 
\DeclareRobustCommand{\DIFdelbeginFL}{\DIFOdelbeginFL \let\includegraphics\DIFdelincludegraphics} 
\DeclareRobustCommand{\DIFdelendFL}{\DIFOaddendFL \let\includegraphics\DIFOincludegraphics} 

\begin{document}

\pagestyle{fancy}
\rhead{\includegraphics[width=2.5cm]{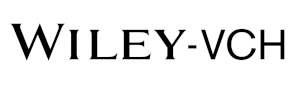}}

\title{Optically trapped exciton-polariton condensates in a perovskite microcavity}

\maketitle

\author{Maciej Zaremba$^{\dag}$}
\author{Mateusz Kędziora$^{\dag}$}
\author{Laura Stańco}
\author{Krzysztof Piskorski}
\author{Kamil Kosiel}
\author{Anna Szerling}
\author{Rafał Mazur}
\author{Wiktor Piecek}
\author{Andrzej Opala}
\author{Helgi Sigur{\dh}sson*}
\author{Barbara Piętka*}

\begin{affiliations}
M. Zaremba, M. Kędziora\\
Institute of Experimental Physics, Faculty of Physics, University of Warsaw, ul.~Pasteura 5, PL-02-093 Warsaw, Poland\\
$^{\dag}$These authors contributed equally.

L. Stańco, K. Piskorski, Prof. K. Kosiel, Prof. A. Szerling\\
Łukasiewicz Research Network - Institute of Microelectronics and Photonics, Warsaw, Poland\\

Dr. R. Mazur, Dr. W. Piecek\\
Institute of Applied Physics, Military University of Technology, Warsaw, Poland\\

Dr. A. Opala\\
Institute of Experimental Physics, Faculty of Physics, University of Warsaw, ul.~Pasteura 5, PL-02-093 Warsaw, Poland\\
Institute of Physics, Polish Academy of Sciences, al.~Lotników 32/46, PL-02-668 Warsaw, Poland\\

Dr. H. Sigur{\dh}sson\\
Institute of Experimental Physics, Faculty of Physics, University of Warsaw, ul.~Pasteura 5, PL-02-093 Warsaw, Poland\\
Science Institute, University of Iceland, Dunhagi 3, IS-107 Reykjavik, Iceland\\
Email: Helgi.Sigurdsson@fuw.edu.pl\\

Prof. B. Piętka\\
Institute of Experimental Physics, Faculty of Physics, University of Warsaw, ul.~Pasteura 5, PL-02-093 Warsaw, Poland\\
Email: Barbara.Pietka@fuw.edu.pl

\end{affiliations}


\begin{abstract}

We demonstrate room temperature optical trapping and generation of high-order angular harmonics in exciton-polariton condensates in a monocrystalline CsPbBr$_{3}$ perovskite-filled microcavity. Using an annular nonresonant excitation profile focused onto the perovskite, we observed power-driven switching between different transverse modes of the optically induced trap. We explore the interplay between the perovskite crystal dimensions and the optical trap diameter that allows the condensate to transition from whispering gallery-like petal shapes to ripple states. Our results underline the feasibility in creating high-order quantum states in perovskite polariton condensates for reconfigurable and structured room temperature nonlinear lasing.
\end{abstract}

\section{Introduction}
Exciton-polariton bosonic quasiparticles (hereafter {\it polaritons}) arising from the strong coupling between cavity photons and excitons, can transition into a power-driven nonequilibrium Bose-Einstein condensate displaying unconventional superfluidity and coherent light emission~\cite{Amo2009, Sanvitto2010, Byrnes2014, Juggins2018}. Quantum fluids of polaritons possess inherently strong optical nonlinearities due to the large Coulomb interaction strength of their exciton constituent which enables a unique optical trapping mechanism for polariton condensates~\cite{Askitopoulos2013, Cristofolini2013}. There, an incoherent (nonresonant) optical pump, with an annular shaped transverse beam profile, induces a polariton potential barrier at the maxima of the focused field. The effective potential originates from a photoexcited background of incoherent excitons, which drive polaritons into the ring center, facilitating their stimulated scattering and condensation. Because the optically trapped condensate is mostly separated from the surrounding pumped hot region, its coherence times can increase by a factor of $10^3$ compared to the polariton lifetime~\cite{Orfanakis2021, Sigurdsson2022, Baryshev_PRL2022}.

To date, optical trapping of polaritons has mainly been studied in inorganic cavities (e.g. CdTe and GaAs quantum wells) which, although featuring good stability under repeated excitation, require cryogenic conditions (few Kelvin) due to the small exciton binding energies. In these systems, a wide range of phenomena has been demonstrated in single traps, such as condensation into high-order modes~\cite{Manni2011, Tosi_NatPhys2012, Cristofolini2013, Dreismann2014, Askitopoulos2015, Sun2018, Nalitov2019}, spontaneous Larmor precession of the polariton pseudospin~\cite{Orfanakis2021, Sigurdsson2022, Baryshev_PRL2022}, Rabi cycling vortex clusters~\cite{Sitnik2022}, polarization bifurcations~\cite{Ohadi2015} and bistability~\cite{Pickup2018}, vorticity~\cite{Dall2014, Ma2020}, tunable macroscopic Bloch vector precession between two trap modes and associated Lotka-Volterra dynamics~\cite{Topfer2020}, exceptional points~\cite{Gao2015}, measurement of the polariton interaction strength~\cite{Estrecho2019}, quantum degeneracy~\cite{Sun_PRL2017}, depletion~\cite{Pieczarka2020} and Bogoliubov excitations~\cite{Pieczarka2022B}. More recently, optical traps of broken axial symmetry were rotated at GHz rates to induce preferential vorticity in polariton condensates~\cite{Gnusov2023, Redondo_NanoLett2023, Gnusov_PRB2024} and probe polariton pseudospin resonances through effective magnetic fields~\cite{Gnusov_Optica2024}.

The richness found in trapped polariton condensates, alongside their optical flexibility, has motivated researchers to explore room temperature operation using materials with much stronger exciton binding energies and light-matter Rabi coupling strength~\cite{Guillet_CRP2016, Wei2022}. For this purpose, lead halide perovskites (LHPs) are a promising candidate~\cite{Fieramosca2019, Su2021} with their high optical absorption coefficients~\cite{Su2017}, tunable photon and polariton lasing~\cite{Zhang_NanoLett2021, Lempicka2024}, large polariton nonlinearities~\cite{Fieramosca2019}, and relatively easy fabrication process into various microstructures~\cite{Kedziora2024}. Several important polariton features have already been demonstrated in perovskites such as coherent ballistic transport~\cite{Su_Science2018, Xu_NatComm2023}, long range spin Hall currents~\cite{Liang_NatPhot2024, Shi_NatMat2025}, analogue XY spin simulation~\cite{Tao2022}, optical switches~\cite{Feng_SciAdv2021, Su_SciAdv2021, Lempicka2024}, electrically driven spin currents~\cite{Wang_AdvMat2024}, and superfluidity~\cite{Peng2022}. However, scarce attention has been given to optically trapped polariton condensates in perovskites likely due to sample roughness which can spoil the pump-induced trap. Instead, researchers have relied on accidental sample defects~\cite{Zhai_PRL2023} or irreversibly patterned structures~\cite{Su_NatPhys2020, Wang2021, Polimeno_AdvMat2024} to trap perovskite polariton condensates. These strategies, however, do not possess the in-situ tunability that reprogrammable optical traps offer. Here, we fill in this gap by demonstrating room-temperature optical trapping of exciton-polariton condensates in microcavities containing CsPbBr$_{3}$ monocrystalline perovskite microwires of varying optical trap widths. 

For this purpose, we have developed a fabrication technique to produce a sample containing perovskites in crystalline form, embedded within distributed Bragg reflectors (DBRs). \textbf{Figure~\ref{fig:sample}} illustrates the fabrication process: the perovskite precursor solution was dripped into a PDMS template with predefined diameter and height, positioned on a substrate with dielectric DBRs, monocrystalline CsPbBr$_{3}$ microwires were then grown using microfluidic-assisted crystallization. Following this step, the PDMS template was removed, leaving the perovskite crystals directly standing on the DBR substrate without any additional layers or covers. Finally, the structure was covered with top DBRs using a low-temperature plasma-enhanced chemical vapor deposition (PECVD) technique to achieve microcavity confinement. This unique fabrication technique integrates two advanced technologies critical for forming a refined high Q-factor cavity: the growth of CsPbBr$_{3}$ in monocrystalline form with a precisely defined cavity thickness achieved through the predefined geometry of the PDMS template, and a top DBR mirror deposition method ensuring uniform, dense, and conformal coverage of the perovskite crystal surface.

By tailoring the LHP microwire geometry and power of the optical trap, the condensation of polaritons can be switched among different high-order trapped modes, accompanied by redistribution of spatial densities and superlinear increase in the emission intensities [see \textbf{Figure~\ref{fig:setup}} and \textbf{Figure~\ref{fig:states}}], implying that polariton condensates in this geometry could be exploited for an all-optical multistate switch. For low powers, we observe condensation into extremely high-order angular harmonics (Laguerre-Gaussian), reminiscent of whispering gallery modes, of orbital angular quantum number up to $l=19$. For higher powers we observe collapse of the condensate to successively lower order angular harmonics as stimulated scattering and phonon-mediated energy relaxation become more efficient at higher densities. For especially narrow wires (up to 5~$\mu$m, where the standard width of microwires in the experiment is 15~$\mu$m) we create elongated trapping conditions which induce condensation into planar Hermite-Gaussian states with different mode numbers along the minor and major trap axis, also referred as {\it ripple} states~\cite{Sun2018}. The results underpin the potential to all-optically engineer single-particle quantum states populated by exciton-polariton condensates in LHPs.

\section{Results}

Our monocrystalline CsPbBr$_3$ microwire sample of predefined diameter and height was created by microfluidic assisted crystallization (for more details see Ref.~\cite{Kedziora2024}). The perovskites were first deposited on 6.5 pairs of SiO$_{2}$/TiO$_{2}$ DBRs and after crystallization covered with 10.5 pairs of Si$_{3}$N$_{4}$/SiO$_{2}$ DBRs at low temperature [see \textbf{Figure~\ref{fig:sample}}] to obtain microcavity confinement with a photon stopband centered at 535 nm. The low-temperature layers deposition (about 370~K) ensures that the perovskite structure is not disturbed, which allows to keep low the number of defects, increasing the optical quality of the structure. Even though the fabrication method of the LHP cavity is different, the final structure is similar to that described in Ref.~\cite{Feng_SciAdv2021}.

A picosecond pulsed laser beam of $\lambda = 435.9$~nm was used to nonresonantly pump the sample at large excitation densities. The ring-shaped optical trap was realized using an 1$^\circ$ axicon lens and 1~mm spatial aperture with an obstruction thickness of 850 $\mu$m was used to eliminate residual light inside the trap and provide a ring of small thickness (sub-micrometer in our case) [see Figure~\ref{fig:setup}\textbf{b}]. The profile of the pump can be described approximately by an annular Gaussian function, 
\begin{equation} 
P(r) \sim \exp \left( - \frac{(r - R)^2}{2\sigma^2} \right) + \exp \left( - \frac{(r +R)^2}{2\sigma^2} \right),
    \label{realringpotential}
\end{equation}
of radius $R$ and thickness $\sigma$. Here, $r = |\mathbf{r}|$ where $\mathbf{r}=(x,y)$ is the cavity in-plane coordinate. Importantly, the experimental configuration does not impart any angular momenta on the beam, and the polarization of the excitation laser was vertical, i.e. aligned with the short axis of the wire, and the same polarization was used for the collected signal. When pumped below threshold, the spatial photoluminescence (PL) pattern forms a slightly smaller ring with respect to the pump circumference and has a wider spread indicating exciton reservoir diffusion from the pumped region as seen in \textbf{Figure~\ref{fig:setup}\textbf{c}} [more details in \textbf{Figure S4} in SI]. When pumped above the condensation threshold, the in-plane coherence of the polariton condensate extends to a size defined by the geometry of the optical trap. The energy of the selected state is determined by a balance of gain and loss dictated by the pump~\cite{Nalitov2019, Askitopoulos2015, Cristofolini2013} and sample parameters, including the LHP exciton-photon detuning. 

The LHP microwires host Wannier-Mott excitons with resonance energy centered at $E_{\text{X}} = 2.375$~eV. The sample Q-factor is around $\approx 340$, with a characteristic photon lifetime of $\tau \approx 100$~fs. Due to the large cavity thickness (600 nm) and birefringence, there are two pairs of horizontal (H) and vertical (V) polarization split lower-polariton energy branches within the investigated range of energies [see \textbf{Figure~\ref{fig:setup}}\textbf{d} and \textbf{Figure S5} in SI]. Notably, the central energy of each pair of branches is negatively detuned from the exciton level by approximately $\Delta \approx 55$ meV (upper pair) and $\approx 147$ meV (lower pair). This results in two different effective polariton masses $m_{1,2}$ and different relaxation rates of reservoir excitons into the optical trap modes. For low in-plane wavevectors $k$ the observed polariton branches can be fitted with parabolic dispersion relations with effective mass for the upper: $m_{1} = 3.8 \times 10^{-5} \: m_0$ and for the lower: $m_{2} = 3.4 \times 10^{-5} \: m_0$ pairs of branches, where $m_0$ represents the free electron mass. Both pairs are also split into H and V polarization by amount $\epsilon_{1,2}$ attributed to the birefringence of CsPbBr$_3$ crystals~\cite{Tao2022} ($\epsilon_{1} = 35$ meV, $\epsilon_{2} = 20$ meV),
\begin{equation}
    \begin{split}
        E_{V1,H1} & = \bar{E_1} + \frac{\hbar^2k^2}{2m_1} \pm \frac{\epsilon_1}{2} \\
        E_{V2,H2} & = \bar{E_2} +\frac{\hbar^2k^2}{2m_2} \pm \frac{\epsilon_2}{2},
    \end{split}
    \label{polaritondisspertion}
\end{equation}
where $\bar{E}_{1,2}$ are the energies at $k = 0$ ($\bar{E}_{1} = 1.95$ eV, $\bar{E}_{2} = 2.42$ eV). In \textbf{Figure ~\ref{fig:setup}}\textbf{e}, we present the system input-output emission curve and blueshift of the condensate when crossing the threshold. We observe the condensate dominantly populating the 2V branch with negligible emission observed from the remaining three branches.

The potential amplitude of the pump-induced annular trap is directly proportional to the density of background incoherent reservoir excitons $n_{\text{X}}(r)$ which, due to their much heavier mass compared to the polariton, diffuse slowly and therefore are approximately proportional to the pump profile $n_{\text{X}}(r) \sim P(r)$ as seen in Figure~\ref{fig:setup}\textbf{c}. The effective polariton potential coming from many-body collisions can be written $V(r) = 2 g_0 |X_\mathbf{k}|^2 n_{\text{X}} (r)$, where $g_0>0$ is the exciton-exciton dipole interaction strength in the order of $g_0 \sim 1 \ \mu$eV$\cdot \mu$m$^2$ for typical LHP~\cite{Fieramosca2019, Su2021}, and $|X_\mathbf{k}|^2$ is the exciton Hopfield coefficient of the polariton.

Due to their low effective-mass photonic component, LHP cavity polaritons can coherently extend over macroscopic distances~\cite{Su_Science2018, Peng2022, Xu_NatComm2023}, thus making it possible for them to occupy modes over the entire optical trap. The spatial distributions of these modes can be changed considerably with variations in the excitation power and trap size~\cite{Askitopoulos2015, Sun2018, Topfer2020} and microwire widths. Below the condensation threshold, polaritons are incoherent and primarily reside near the ridge of the annular pump region [see Figure~\ref{fig:setup}\textbf{c}, Figure S4 in SI]. When the excitation power surpasses the condensation threshold, polaritons undergo stimulated scattering into a trap state that provides optimal balance between gain and losses (similar to lasers)~\cite{Nalitov2019}. \textbf{Figures~\ref{fig:states}}\textbf{a-d} show the above threshold cavity PL from an optically trapped condensate in a large LHP microwire (i.e., pump is focused entirely within the wire's extend). Observed condensate density profiles belong to a superposition of opposite rotating angular harmonics~\cite{Manni2011, Dreismann2014, Askitopoulos2015, Sun2018} denoted by an orbital angular quantum index $l = 2,4,15,$ and $19$. Here, $l$ is defined as the eigenvalue of the angular momentum operator $\hat{L}_z = -i \hbar \partial_\varphi$ out of the cavity plane. Scanning different locations along the microwire sample allowed to observe different petal patterns due variable focusing of the pump profile (by changing the distance between the objective and the sample), affecting its thickness $\sigma$. The pump parameters for each sample location studied are given in the caption of Figure~\ref{fig:states}\textbf{a-d}. Corresponding spectrally resolved PL results are shown in \textbf{Figure S2} in the SI. For large pump thickness $\sigma$ [Figures~\ref{fig:states}\textbf{a-b}], the optical trap becomes narrow, thus pushing modes out into the continuum and forcing the condensate to populate lower order modes instead in agreement with previous works~\cite{Askitopoulos2013, Cristofolini2013, Sun2018, Wei2022}. Conversely, for small pump ring thickness [Figures~\ref{fig:states}\textbf{c-d}] the trap becomes wider and only high order angular harmonics have enough kinetic energy to push into the pump gain region to condense~\cite{Dreismann2014, Nalitov2019}. Because of the large wire diameter employed here, our results are fully general to other sample types such as LHP nanoplatelets or disks that can contain the excitation profile.

As mentioned above, the condensate populates the 2V branch and therefore the observed petal structures are determined by its effective mass $m_2$. We stress that H polarized emission was 10-times lower in intensity and thus negligible. For thicker pump rings [Figure~3\textbf{a-b}], lower power densities are needed to reach threshold due smaller trap lateral dimension. For narrow pump rings [Figure 3\textbf{c-d}], larger power density is needed to reach threshold. We note that when the system is pumped strongly out of equilibrium (not shown) the condensate is eventually pushed into the lower energy 2H branch, indicating enhanced relaxation. 

The observed condensate PL patterns can be fitted using the eigenstates $|\psi_{l}\rangle$ of $\hat{L}_z$ belonging to a rotationally symmetric two-dimensional harmonic confining potential:
\begin{equation}
\psi_{l}(r, \varphi) = \frac{\beta}{\sqrt{2 \pi l!}} e^{i l \varphi} (\beta r)^{l} e^{-\beta ^2 r ^2 /2}.
\label{psind}
\end{equation}
Here, $\beta$ is a fitting parameter denoting the strength of the ``harmonic'' confinement coming from the optical trap. The steady state condensate profiles correspond then to:
\begin{equation} \label{eq.rho}
\rho(\mathbf{r}) = |\psi_{-l} + \psi_{l}|^2,
\end{equation}
where $\psi_{-l}$, $\psi_{l}$ demonstrate two propagating polariton condensate eigenfunctions ($\psi_{l}$ clockwise and $\psi_{-l}$ counterclockwise) inside the excitation profile, which are plotted in Figure~\ref{fig:states}\textbf{e-h} showing good agreement. It is essential to highlight the good contrast of the measured petal shaped condensates despite operating at room temperature with broad linewidth excitons. To our knowledge, this result represents the first realization of such high-order polariton condensate states in an optical trap at room temperature. Note, the deformation of the state in Figure~\ref{fig:states}\textbf{b} is connected with proximity of the microwire edge, also visible on the same panel as the scattered PL signal at the wire edge located along \textit{x} = 5~$\mu$m. 

We next studied the case where the pump diameter is bigger than the LHP microwire to explore finite size effects from the wire [see \textbf{Figure~\ref{fig:asymetric}]}. Coincidentally, at low pump powers we observe PL coming from a defect located at $x \approx -2.5$~$\mu$m [see Figure~\ref{fig:asymetric}\textbf{a}]. Hence, the condensate pattern becomes slightly asymmetric, causing an uneven mode distribution near $2.304$~eV and $2.264$~eV. Such defects are a common issue in LHPs (for details on LHP crystal quality, see Ref.~\cite{Kedziora2024}). As the power is increased, we eventually observe dominant emission coming from the polariton condensation populating a Hermite-Gaussian $\psi_{n_x=1,n_y=4}$ mode [see Figure~\ref{fig:asymetric}\textbf{b}] with different standing wave quanta $(n_x,n_y)$ along $x$ and $y$ due to the small width of the wire compared to the pump diameter. The combined in-plane potentials from the finite wire width (confinement along $x$ acting on photons) and the optical trap (confinement along $y$ acting on excitons) reduce the rotational symmetry of the condensate down to $\mathcal{C}_2$ rotation ($180^\circ$) which means orbital states are no longer good eigenmodes and become replaced by Hermite-Gauss modes. The energy-resolved real-space PL shown in Figure~\ref{fig:asymetric}\textbf{c} was collected along the $y$-axis marked by black dotted line in Figure~\ref{fig:asymetric}\textbf{b}. There we can see two major parts of the spectra -- the PL around 2.30~eV is connected with sample defects and wire's edges -- while the much stronger and narrow lower energy PL around 2.28~eV corresponds to the trapped polariton condensate. Figure~\ref{fig:asymetric}\textbf{d} shows energy resolved cross-sections of the two states visible around 2.278 eV showing different parity. 

By introducing a slight power density gradient to the laser beam profile breaking axial symmetry~\cite{Cristofolini2013, Askitopoulos2015, Sun2018}, high-order standing wave condensates can be tailored across the microwire, resulting in ripple condensates shown in \textbf{Figure~\ref{fig:tunable}}\textbf{a} and~\ref{fig:tunable}\textbf{b}. Here, we observe a collapse of the condensate to a lower order mode when the pump power is increased from 1.1 to 1.4$P_\text{th}$ due to increased influence of incoherent reservoir excitons assisting with the thermalization of polaritons. That is, polariton-polariton and polariton-reservoir interactions increase around the pumped region and subsequent pair-scattering into the available lower energy states of the optical trap is enhanced~\cite{Wertz_PRL2012}. Therefore, at sufficient powers, the condensate starts populating a new state at lower energies with lower losses, as reported in ~\cite{Topfer2020, Wei2022}. We investigated this quantum-state selectivity of the condensate by performing a power scan, in units of threshold power $P_\text{th}$, where we observe the state $\psi_{n_x=10,n_y=0}$ [Figure~\ref{fig:tunable}\textbf{a} at $1.1 P_{\text{th}}$] smoothly transition into $\psi_{n_x=9,n_y=0}$ [Figure~\ref{fig:tunable}\textbf{b} at $1.4 P_{\text{th}}$]. For additional measurements taken far above $P_{\text{th}}$ refer to \textbf{Figure S3} in the SI. The corresponding normalized cross sections of the condensate spatial density profiles are shown in Figure~\ref{fig:tunable}\textbf{c} and~\ref{fig:tunable}\textbf{d}. The power scan is shown in Figure~\ref{fig:tunable}\textbf{e} where the number of nodes (real space interference fringes) reduces gradually at higher powers. This pronounced mode switching behavior, clearly resolvable in our system, highlights an inherent advantage of higher-order quantum states, where distinct transitions become more accessible and are difficult to quantify in lower-order systems. Our findings are consistent with recent experiments in low-temperature III-V microcavity systems~\cite{Topfer2020} and organic cavities~\cite{Wei2022}.

\section{Discussion and Conclusions}
We have demonstrated optical trapping and quantum state tailoring for exciton-polariton condensates at room temperature in CsPbBr$_{3}$ perovskite microcavities. Our nonresonant optical pump forms an annular ring, which provides efficient transverse confinement of polaritons through their strong repulsive Coulomb interactions with a photogenerated background of hot excitons. Because of the driven-dissipative nature of the system, the observed stable condensate profiles correspond to high-order quantum states, including whispering-gallery like modes (petals) and bouncing-ball modes (ripples). The all-optical trapping aspect facilitates easy switching between these states and possible generation of complex superpositions of different states competing over the pump gain~\cite{Topfer2020}. The patterns resemble those found in multimodal vertical-cavity surface-emitting lasers~\cite{Li2012} and in low-temperature inorganic polariton systems~\cite{Askitopoulos2015, Dreismann2014, Sun2018}. Here, however, our perovskite system combines the advantages of all-optical reconfigurability for generation of coherent and highly nonlinear structured light with room temperature operation. Their solution-processable, reconfigurable nature supports scalable photonic circuits~\cite{Kedziora2024} making them ideal for integration with other materials such as silicon, GaN, Si$_3$N$_4$ or liquid crystals to explore synthetic photonic spin-orbit fields~\cite{Lempicka2022}. Our system is also promising for future investigation into polariton condensation into trapped states defined by high order angular momenta~\cite{Zhai_PRL2023, Wang_NSR2022, Shang_AOM2024, Zhang2025, Dong_ACS2025}. For example, in cold atom systems, vortex formation is central to phenomena like quantum turbulence and superfluidity~\cite{Matthews1999, Fetter2009}. In condensed matter physics, vortex states emerge in systems like superconductors and superfluids, and these excitations have become important for exploring quantum phase transitions and exotic quantum states~\cite{Nernst1915}. Additionally, vortex-like excitations are explored in nonlinear optics, where they are used to study the dynamics of optical beams and wavefronts, as well as their potential in quantum information processing~\cite{Grosso2007, Lehur2012}. By drawing parallels between these systems and our perovskite-based polariton condensates, we open new directions for investigating high-order orbital angular momentum dynamics in nonlinear optical systems.

Given previous demonstrations of long-distance polaritonic propagation in microwires and waveguides~\cite{Su_Science2018, Xu_NatComm2023, Peng2022} and susceptibility to engineered photonic potentials~\cite{Nguyen2013}, there are perspectives to combine structured optical excitation with aforementioned methods to explore new directions to control polaritons at room temperature. Already there have been demonstrations of perovskite polaritons for optical switching in either amplitude~\cite{Feng_SciAdv2021, Su_SciAdv2021} or polarization~\cite{Lempicka2024, Shi_NatMat2025}. Another avenue involves reprogrammable condensate lattices with tunable inter-site coupling strengths through barrier height control, facilitating applications such as many-body quantum circuitry, neuromorphic and analogue optical computing and quantum simulators~\cite{Opala_23, Kavokin2022}. Unconventional pump geometries, such as billiards~\cite{Gao2015} and stadiums~\cite{McDonald1988}, can be implemented using spatial light modulator to explore polariton chaos. Despite significant progress, challenges remain in stability, scalability, and device integration.

Another optical trapping technique worth exploring in the context of perovskites is subwavelength grated photonic structures which host negative-mass bound in the continuum polaritons that experience strong confinement under localized Gaussian pumping, instead of ring shaped pumping, with very long lifetimes~\cite{Gianfrate_NatPhys2024}.

\medskip
\textbf{Acknowledgements} \par 
M.Z. and B.P. acknowledge the European Union EIC Pathfinder Open project “Polariton Neuromorphic Accelerator” (PolArt, Id: 101130304). M.K., R.M., W.P. acknowledge the National Science Center, Poland, project No. 2022/47/B/ST3/02411. L.S., K.P., K.K. and A.S. acknowledge the support by the statutory funds of the Łukasiewicz Research Network – Institute of Microelectronics and Photonics. H.S. acknowledges the Icelandic Research Fund (Rann\'{i}s), grant No. 239552-051. M.Z. and H.S. acknowledge the project No. 2022/45/P/ST3/00467 co-funded by the Polish National Science Center and the European Union Framework Programme for Research and Innovation Horizon 2020 under the Marie Skłodowska-Curie grant agreement No. 945339. A. O. acknowledges the National Science Center, Poland, project No. 2024/52/C/ST3/00324.

\medskip
\textbf{Competing interest} \par 
\noindent The authors declare no competing interests.

\medskip
\medskip

\bibliographystyle{MSP}
\bibliography{apssamp}
\begin{figure}
\centering
\includegraphics[width=0.8\linewidth]{Fig1.png}
\caption{Sample preparation process. \label{fig:sample}\textbf{a} Dripping the prepared solution in PDMS template of predefined diameter and height prepared on the 6.5 pairs of SiO$_{2}$/TiO$_{2}$ DBRs' substrate. \textbf{b} Monocrystalline CsPbBr$_3$ microwire growth by microfluidic-assisted crystallization. \textbf{c} The low-temperature PECVD technique covering with 10.5 pairs of Si$_{3}$N$_{4}$/SiO$_{2}$ DBRs to obtain microcavity confinement with a photon stopband centered at 535 nm.}
\end{figure}

\begin{figure}
\centering
\includegraphics[width=\linewidth]{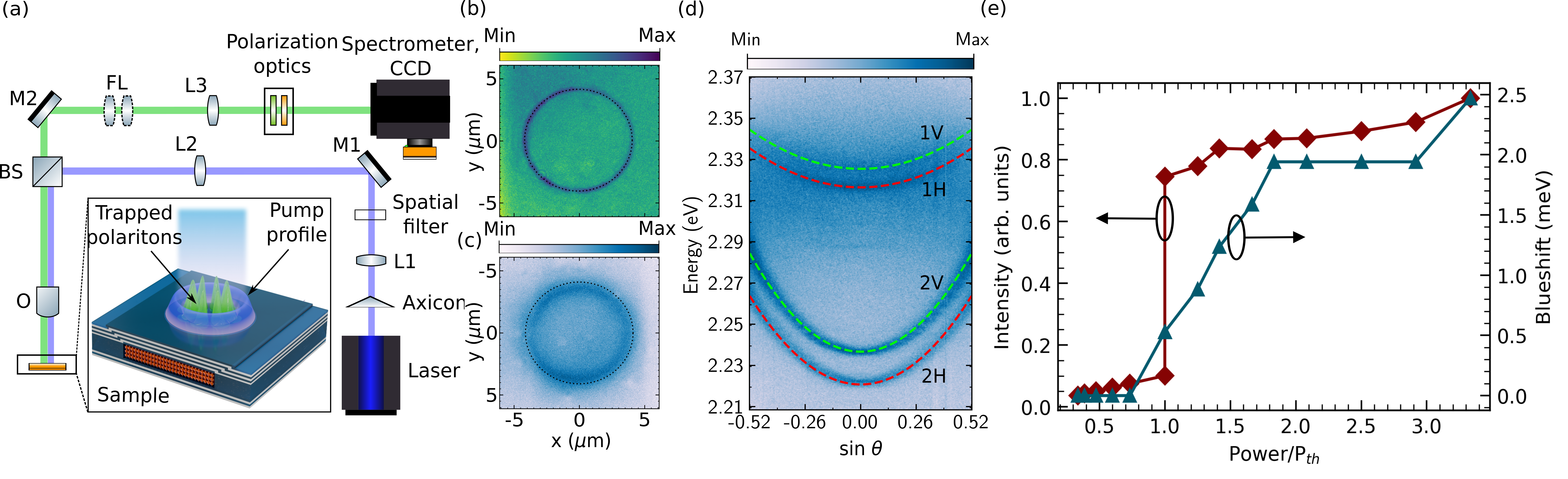}
\caption{Experiment preparation and principal measurments. \label{fig:setup}\textbf{a} A scheme of the experimental setup composed of beamsplitter (BS), objective (O), mirrors (M), lenses (L) and Fourier lenses (FL). The inset illustrates schematically the optical trap and the sample. Panel \textbf{b} shows the prepared annular laser beam profile focused on the sample and panel \textbf{c} the corresponding luminescence below threshold. The dashed circular lines help to demonstrate, as a guide-to-the-eyes, that the hot exciton reservoir is created within the pump profile. Panel \textbf{d} shows the angle resolved PL of the system below threshold, revealing two pairs of H-V (red and green curve) split polariton branches. The dashed lines are parabolic fits to the data using Eqs.~\eqref{polaritondisspertion}. The data is presented in respect to the $\sin \theta$ (where 
$\theta$ is the angle at which the photoluminescence signal was collected in the experiment and $k=E \sin \theta / hc$, where $E$ is the polariton energy). Panel \textbf{e} demonstrates the power-dependent integrated emission intensity (red curve, arrow to the left Y-axis) and corresponding blueshift of the condensate emission line (blue curve, arrow to the right Y-axis) in units of condensation threshold power $P_\text{th}$.}
\end{figure}

\begin{figure}
\includegraphics[width=\linewidth]{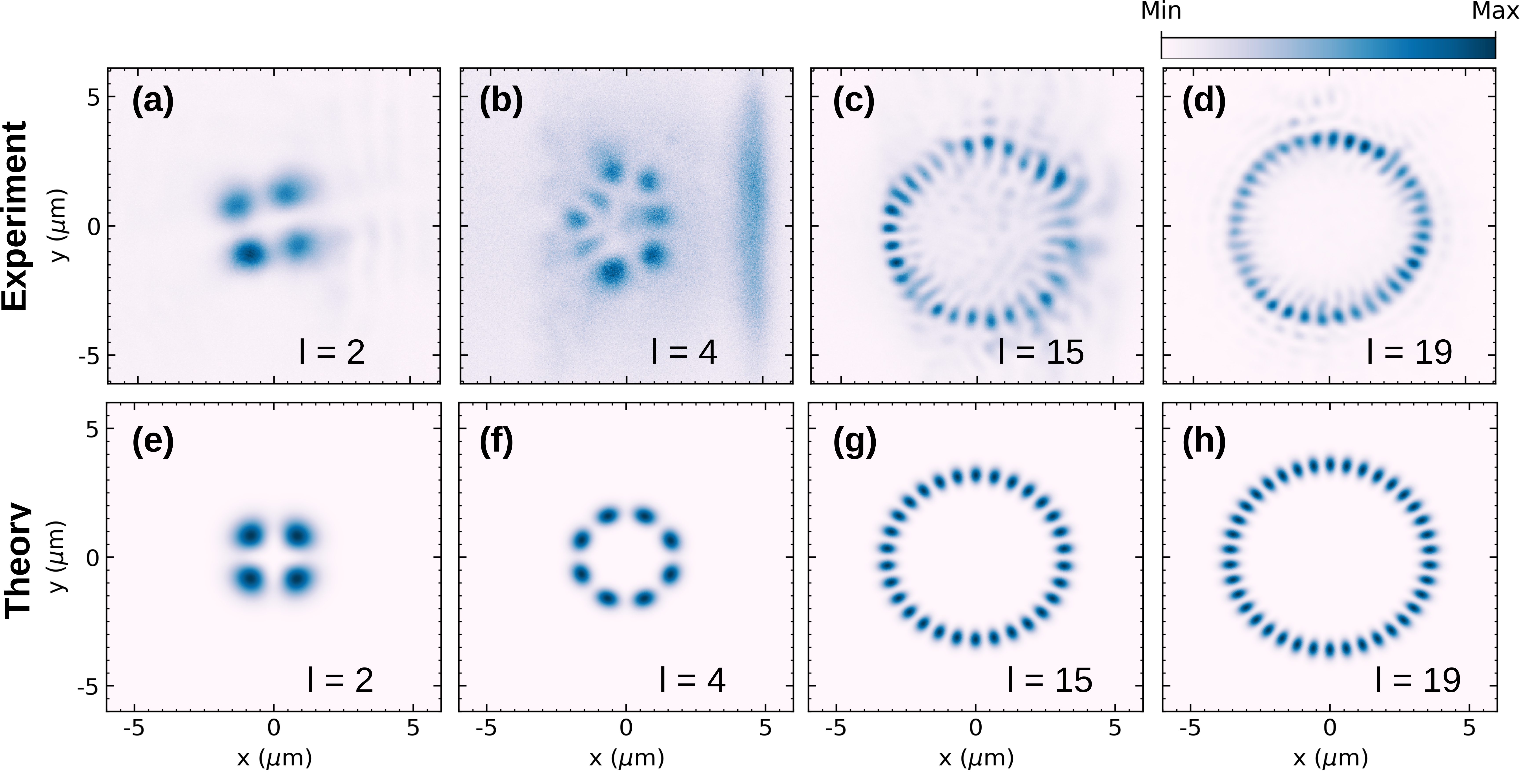}
\caption{\label{fig:states} Optically trapped polariton condensate harmonics of varying angular momentum $l$. All results are resolved in V-polarization and pumped $\approx10\%$ above condensation threshold. The pump is centered on the microwire which is much broader. Panels \textbf{a}-\textbf{d} show the experiment, and \textbf{e}-\textbf{h} show corresponding theoretical profiles from~\eqref{eq.rho}. Condensates petals formed by ring-shaped exciton reservoirs of the spatial distribution described by~\eqref{realringpotential} with radius $R = 3.5$ $\mu$m and different thickness parameters $\sigma$ for each panel, i.e. \textbf{a}: $\sigma \approx 1.5$ $\mu m$, \textbf{b}: $\sigma = 1.3$ $\mu$m; \textbf{c}: $\sigma = 0.9$ $\mu$m; \textbf{d}: $\sigma = 0.6$ $\mu$m.}
\end{figure}

\begin{figure}
\centering
\includegraphics[width=0.7\linewidth]{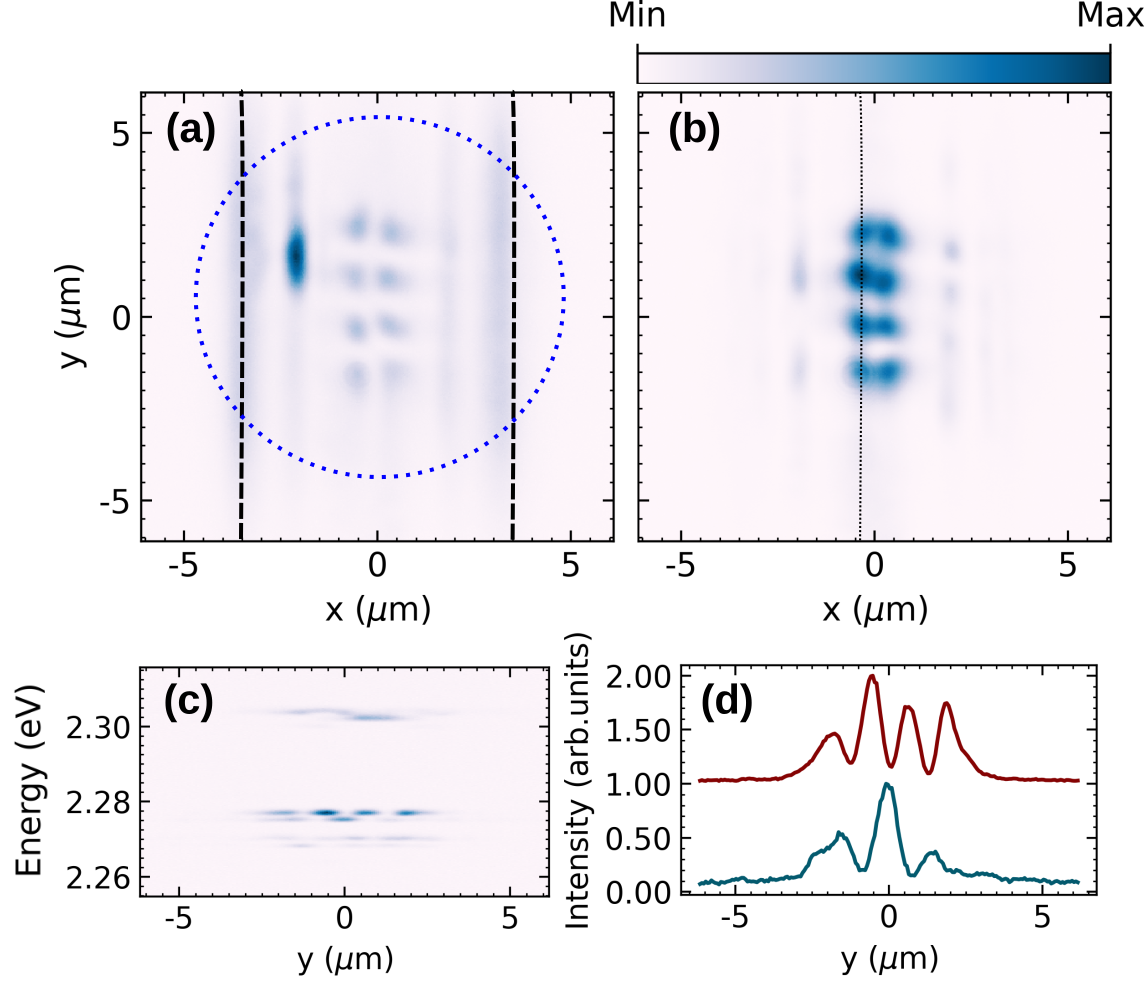}
\caption{\label{fig:asymetric} Panel (a) shows the emission at threshold for a configuration in which the trap (blue dots) is larger than the wire diameter (black dashed lines). The bright emission around \( x = -2.5 \,\mu\text{m} \) is attributed to a sample defect. Panel (b) depicts the emission above (20\%) threshold, and panel (c) presents real-space spectra acquired under the same conditions as in (b), with the locations indicated by black lines. The eight trapped nodes of the condensate in panel (a) are associated with modes in the spectrum that are symmetrically distributed in the dispersion [see panel 4(c)] in the vicinity of energies \( 2.278 \,\text{eV} \) and \( 2.268 \,\text{eV} \), whereas the remaining observed modes, which are asymmetrically distributed around \( 2.304 \,\text{eV} \) and \( 2.264 \,\text{eV} \), are related to condensation at the defect visible in panel (a) at \( x = -2.5 \,\mu\text{m} \) and \( y = 1.5 \,\mu\text{m} \). Panel (d) shows a cross section through two states visible around \( 2.278 \,\text{eV} \).}
\end{figure}

\begin{figure}
\centering
\includegraphics[width=0.6\linewidth]{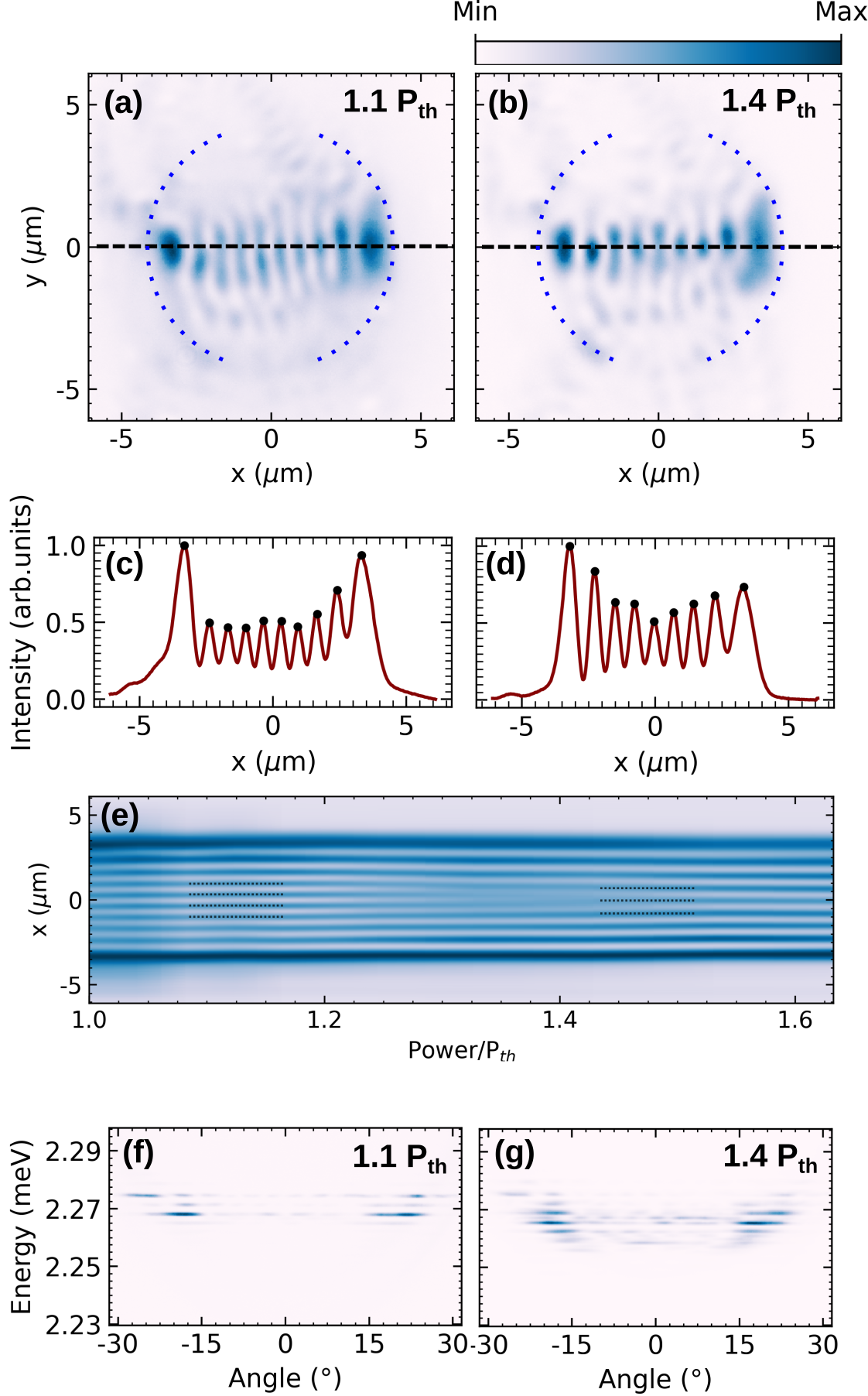}
\caption{\label{fig:tunable} Tunability of mode number by pump power. Panels \textbf{a} and \textbf{b} shows real space images of trapped condensate at 1.1 and 1.4 P$_{th}$ using an asymmetric trap profile as indicated with blue dashed circle, respectively, \textbf{c} and \textbf{d} are cross-sections of \textbf{a} and \textbf{b} at $y = 0$ as indicated by dashed lines.  Panel \textbf{e} show power-scan of the ripples number and black dotted lines are to guide the eye. The angle-resolved spectra of the condensates for pumping powers as in \textbf{a} and \textbf{b} are visible on \textbf{f} and \textbf{g}. Condensate ripples formed by ring-shape exciton reservoir described by~\eqref{realringpotential} with parameters: $R = 3.5$ $\mu$m, $\sigma = 1.5$ $\mu$m}
\end{figure}

\begin{figure}
\textbf{Table of Contents}\\
\medskip
  \includegraphics{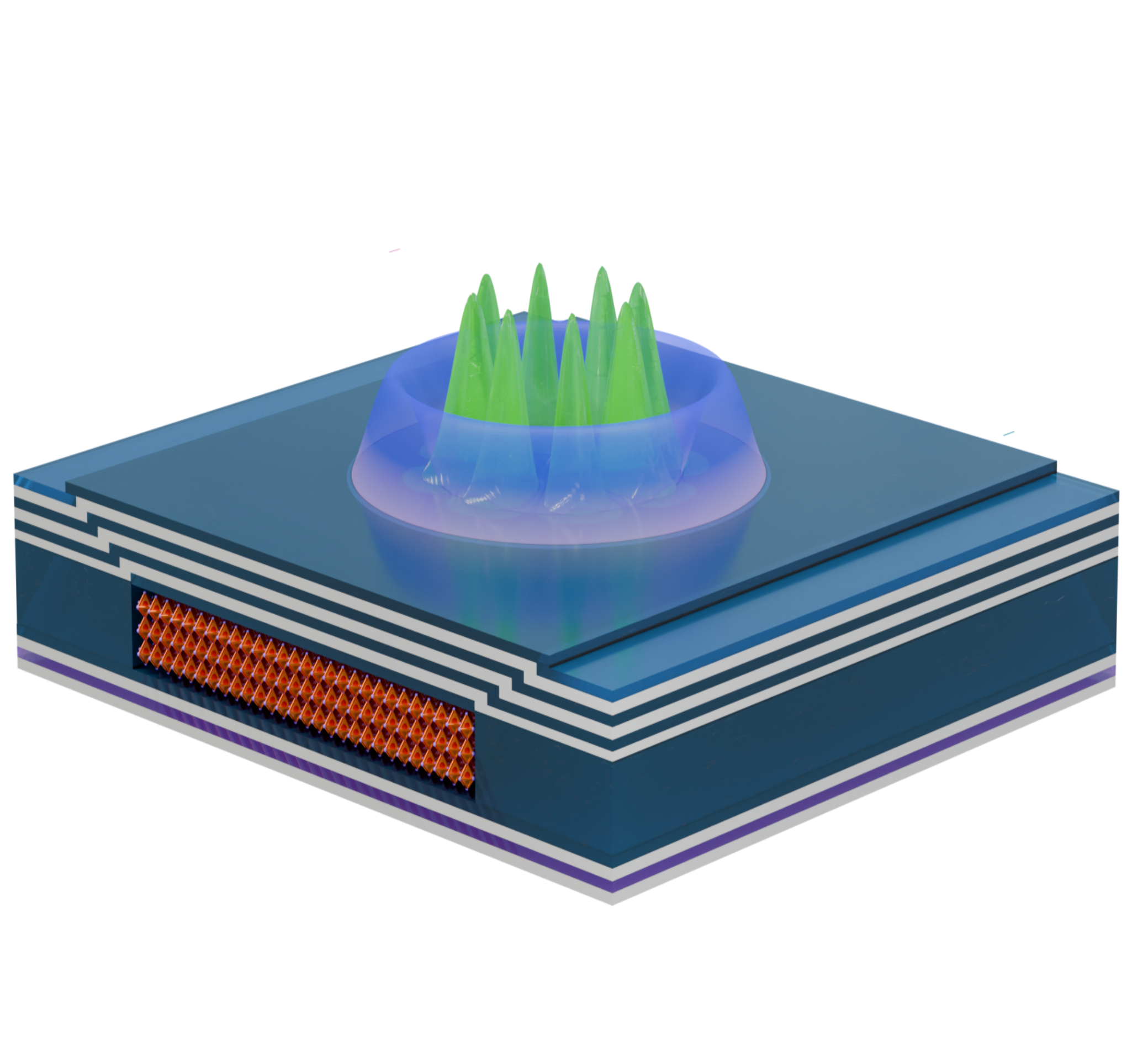}
  \medskip
  \caption*{Researchers report room-temperature optical trapping and the generation of high-order angular harmonics in exciton-polariton condensates within a CsPbBr$_{3}$ perovskite-filled microcavity. By varying the optical trap size and crystal dimensions, they demonstrate transitions between whispering gallery-like petal modes and ripple states, paving the way for structured, reconfigurable nonlinear lasing at room temperature.}
\end{figure}

\end{document}